\documentclass{jfm}
\usepackage{natbib}
\usepackage{epsfig}
\usepackage{AMSsymb}

\def\x{{\mbox{\boldmath$x$}}}
\def\u{{\mbox{\boldmath$u$}}}

\def\R{{\mbox{\boldmath$R$}}}

\def\unitR{{\mbox{\boldmath$\hat R$}}}

\def\begineq{\begin{equation}}
\def\endeq{\end{equation}}
\def\be{\begin{equation}}
\def\ee{\end{equation}}
\renewcommand{\S}{{\cal S}}
\newcommand{\bm}[1]{\mbox{\boldmath $#1$}}
\newcommand{\bu}{{\mbox{\bm u}}}
\newcommand{\bv}{{\mbox{\bm v}}}

\newcommand{\bR}{{\mbox{\bm R}}}

\newcommand{\bRhat}{\mbox{\bm {\hat R}}}
\newcommand{\bx}{\mbox{\bm x}}

\newcommand{\lp}{\left(}
\newcommand{\rp}{\right)}
\newcommand{\la}{\left\langle}
\newcommand{\ra}{\right\rangle}

\title[Probing structures in channel flow through SO(3) and SO(2) decomposition]{Probing structures in channel flow through SO(3) and SO(2) decomposition}
\author[Luca Biferale, Detlef Lohse, Irene M. Mazzitelli and  Federico Toschi]
{Luca Biferale$^{1}$, Detlef Lohse$^{2}$, Irene M. Mazzitelli$^{2}$, and\\  Federico Toschi$^{2}$}

\affiliation{$^1$Department of Physics and INFM,\\
University Tor Vergata, Via della Ricerca Scientifica 1, 00133 Rome, Italy\\[\affilskip]
$^2$Department of Applied Physics  and J.\ M.\ Burgers Centre for Fluid Dynamics\\
University of Twente, 7500 AE Enschede, Netherlands}

\pubyear{2000}
\volume{538}
\pagerange{119--126}
\date{?? and in revised form ??}
\begin{document}
\maketitle
%
\begin{abstract}
SO(3) and SO(2) decompositions of numerical channel flow turbulence
are performed. The decompositions are
 used to probe, characterize, and quantify 
anisotropic structures in the flow. 
Close to the wall the anisotropic  modes  are dominant
and reveal the flow structures. 
The SO(3) decomposition 
does not converge for large scales as expected. However, in the shear
buffer layer 
 it also does not converge for small scales, reflecting the lack
of small scales isotropization in that part of the channel flow. 
\end{abstract}


\section{Introduction}
Kolmogorov theory of fully developed turbulence has defined the main
stream  for almost all theoretical and applied turbulence investigations since
it appeared in the 1941 (\cite{kol41})\\
Kolmogorov made two important statements  (i) there exists a  
tendency of any turbulent flow
towards isotropization and 
homogenization of  small scales fluctuations, (ii) there exists   an inertial range
of scales where the 'almost' isotropic and homogeneous turbulent fluctuations
are characterized by a power law spectrum with a universal $-5/3$ slope.
Both statements  are connected and only partially correct. \\
First, it is well established
experimentally and numerically, \cite{fri95},
that already in the ideal isotropic and homogeneous 
high-Reynolds numbers limit 
turbulent fluctuations cannot be characterized by only one single set of spectrum exponent,
i.e. velocity fluctuations  are strongly intermittent. Intermittency is 
the way to summarize
the fact that  the probability  density
of velocity increments, $\delta_R u = \lp \bu(\bx + \bR) - \bu(\bx) \rp\cdot \bRhat$,  over a distance
 $R$ cannot be rescaled 
by using only one single scaling
exponents for all $R$, see for example \cite{fri95}.  Second, in almost all 
relevant applied situation one is interested in those ranges of scale where
turbulence statistics is neither homogeneous nor isotropic. \\ In the following, 
as an example, we will discuss in detail the important case of channel flows.
Recent experimental, \cite{gar98},
and numerical investigations, \cite{pum96,pum95}, have shown that the
tendency towards the  isotropization
of small scale statistics of  shear-flows is much slower
than any   dimensional prediction even at very large
Reynolds numbers; in contrast to what is predicted by the 
 Kolmogorov 1941 theory, some observables
like the skewness of velocity gradients, exhibit  persistence 
 of anisotropies.  The two above issues  are 
connected. One cannot focus on the issue
of intermittency in  high-Reynolds number  homogeneous
and isotropic statistics  without first having  a systematic control on 
the possible slowly decaying anisotropic effects always present in all
numerical or experimental investigations. Similarly, the understanding
of complex non-homogeneous and anisotropic flows cannot avoid the problem of
intermittent isotropic and anisotropic fluctuations. \\

Despite the many systematic theoretical attempts to attack
intermittency  in isotropic and homogeneous turbulence, 
the problem is mainly  unsolved
except for a class of toy-cases considering  the passive advection
by Gaussian and white in time velocity fields of scalar, 
\cite{gaw95,cher95}, or vector quantities, \cite{ver96,lan99,ara00}.
 Nevertheless,
a lot of phenomenological and analytical progresses have been
done by applying 
standard statistical closure techniques, 
\cite{kra72,les87,lvo97a},
or more recent phenomenological tools borrowed from 
dynamical system theory like the fractal and
multi-fractals description of  the energy transfer and of the energy 
dissipation rate, \cite*{ben84,par85,fri95,bohr,ben98,gro00b}.\\
Strangely enough, only very recently, \cite{gro98b,ara98,ara99,ara99b},
similar statistical 
attempts have been transposed to the understanding of 'non-ideal' turbulence,
i.e., turbulence
in all those situations when anisotropy and non-homogeneities play
an important role in the turbulent production  and dissipation. This paper
is meant to partially  fill the gap between the quantitative systematic 
methodology
used in 'ideal' homogeneous and isotropic  turbulence and the qualitative,
ad-hoc, description in terms of 'structures' used in the 'non-ideal'
wall bounded flows. We  show how  the decomposition
of statistically stable observable, like moments of velocity increments, in terms
of the irreducible representations of the group of rotations
in three dimensions, SO(3), and two dimensions, SO(2), allows a quantitative
systematic characterization of the isotropic and anisotropic fluctuations.
Moreover, a connection between the projections on different eigenvectors of
the rotational groups and   structures
 like 'hairpins' and 'streaks' is possible. \\
Structures called 'streaks' 
have been thought to be the main signatures of wall bounded
flows in  the  viscous sub-layer since the pioneering
works of \cite{kline} who observed  the existence of 
extremely well organized
motions made  of region of low and high speed fluid, elongated downstream and
alternating in the span-wise direction. Later, in \cite{kim71},
'streaks' were reported to be the dynamical responsible of turbulent production
in the viscous sub-layer. Similarly, 'hairpins' have been the main persistent
structures observed experimentally, \cite{head,wallace}, 
and numerically, \cite{mk85,km86}, 
outside the viscous layer, in the turbulent boundary layer.   By mean of a
conditional sampling, \cite{km86} were able to show that these
'hairpin' shaped structures are associated with high Reynolds-shear stress
and give a  significant contribution to 
turbulent production in the logarithmic layer.\\
More recently, \cite*{tos99,fede_pof},
 started a first systematic investigation of the intermittent properties of
 velocity increments parallel to the wall
as a function of the distance from the wall in a channel flow simulation.  
In this case a clear transition between the bulk physics 
and the wall physics was recognized in terms of two different set of
intermittent exponents characterizing velocity fluctuations at  the center
and close to the channel walls. Still a firm quantitative understanding
of how much these intermittent quantifiers can be connected to
the presence of persistent structures is lacking. 
 For instance, in  \cite{fede_pof} the different
behavior of velocity 
fluctuations in the buffer layer 
 was explained as a breaking of the 
Kolmogorov refined hypothesis linking energy dissipation to inertial
velocity fluctuations, i.e. an effect due to the different production
and dissipation mechanism caused by the presence of strong
shear effects close to the walls. Clearly, such a kind of issue
can only be addressed  by using systematic tools which are able to 
quantify the degree of anisotropy and coherency at difference scales and at different
spatial locations in the flow.\\
In this paper we propose to use the exact decompositions of the correlation
functions in terms of the irreducible representations of the rotational group
SO(3) (in the bulk of the flow)  and 
in terms of the irreducible representation of the rotational group in two dimensions,
SO(2), (close to the walls) in order to quantify in a systematic way the
relative and absolute degree of anisotropy of velocity fluctuations. Furthermore, we show how a careful analysis of the data allows also for a connection
between some coefficients of the decompositions and the more common
'structures' observed by simple flow visualization. 
We  show how the SO(3) decomposition, being connected to the exact invariance
under rotations of the inertial and diffusive terms of the Navier-Stokes
equations, is able, where applicable, to exactly disentangle universal 
scaling properties of the isotropic sectors from the more complex behavior
in the anisotropic sectors. We also show how the SO(2) decompositions in planes
parallel to the walls, is a useful analyzing tool in order to 
quantify the relative change of planar 
anisotropy by approaching the boundaries.

The paper is organized as follows: In section 2 we review the main 
theoretical consideration about the importance of the SO(3) decomposition in
the Navier-Stokes eqs. In section 3 we present a systematic analysis
of the SO(3) decomposition in a numerical channel flow data base. We discuss
the results with particular emphasis on the universality issue, i.e.
independence from the large scale effects, and on how one can use
such a decomposition to quantify the relative importance of structures
like 'hairpin' in the bulk of the flow. In section 4 
we review the main findings which pushed us to apply the SO(2)
decompositions in planes well inside the buffer layer, i.e. where
the SO(3) decomposition cannot  be applied due to the presence of
the rigid walls,  and we show how the SO(2) analysis allows us
to clearly distinguish the existence of 'streaks' like structures
in a statistical sense.
Section 5 is left to comments and conclusions.

\section{SO(3) decomposition}
SO(3) -- rotational invariance -- is one of the basic symmetries of the
Navier-Stokes equations. However, it is broken by the boundary conditions
or by the driving force of the flow, both of which introduce anisotropy
and also inhomogeneities.
For the sake of completeness let us start, as an example, with the 
the SO(3) decomposition of the 2nd order most general velocity  tensor
depending only on one spatial increment $\R$:
\be
C_{\alpha \beta}(\x ,\R) =
\left< 
(u_\alpha (\x +\R ) - u_\alpha (\x ) 
(u_\beta (\x +\R ) - u_\beta (\x ) 
\right> 
\label{eq1}
\ee
It is easy to realize, \cite{ara99b}, 
 that this observable can be decomposed in terms
of the irreducible representations of the three dimensional rotational group 
which form a complete basis in the space of smooth second order tensors
depending on one vector $\R$:
\be
C_{\alpha  \beta} (\x , \R) = \sum_{qjm} a_{q,jm} (\x , R)
B_{\alpha \beta}^{q,jm} (\unitR),
\label{eq2}
\ee
 The notation in (\ref{eq2}) is borrowed
from the quantum mechanical 
analogue, i.e. $j=0,1,....$ labels the eigenvalues of 
the modulus of the total angular momentum, ${\bf L^2}$; $m=-j,..,+j$
 labels the eigenvalues
of the projection of the total angular momentum on one 
direction, say $\hat y$;
 $q$ labels the different irreducible representations with a given $j$;
and $B_{\alpha \beta}^{q,jm} (\unitR)$ are the eigenfunction of the rotational
group in the space of second order smooth tensors. 
For example for the fully isotropic sector, $j=0$, we have only $m=0$ and
a simple calculation shows that there are only two independent irreducible 
representations in the isotropic sectors, i.e. the well known result,
 \cite{my75}, that  we need   only two independent eigenfunction 
in order to describe any second order
isotropic tensor. These two eigenfunction can be taken
to be: $$B_{\alpha \beta}^{1,00} (\unitR) = \delta_{\alpha,\beta}; \;\;\;\; 
B_{\alpha \beta}^{2,00} (\unitR) = 
{\bf \hat{R}}_{\alpha} {\bf \hat{R}}_{\beta}$$ and therefore
the decomposition (\ref{eq2}) in the isotropic sector
 assumes the familiar form:
\be
C_{\alpha  \beta} (\x , \R) = a_{1,00}(\x,R) \delta_{\alpha,\beta}+
a_{2,00}(\x,R) \unitR_{\alpha}\unitR_{\beta}
\ee
In appendix A we list the   complete set of $B_{\alpha \beta}^{q,jm}$
for the case of second order tensors. For higher tensor ranks 
we refer to \cite{ara99b}.
 The main physical information is of course hidden in the dependence of the  
coefficients $a_{q,jm} (\x , R)$ on the spatial location, $\x$, and on 
the analyzed scale, $R$. 
We aim at  using the decomposition (\ref{eq2}) as a filter able
to exactly disentangle different anisotropic effects as a function
of the spatial location and of the analyzed scale.
In previous studies, the main interest was focused on the theoretical
issue of the existence of scaling 
behavior for the coefficients $a_{q,jm} (\x , R)$
and on its possible dynamical explanation in terms of the 'foliation'
of the Navier-Stokes eqs in different $j$ sectors, \cite{ara98,ara99,ara99b}.
The typical questions addressed were whether coefficients
belonging to different $j$ sectors have different scaling behavior (if any)
and, in the case, which kind of dimensional estimate for scaling
exponents in the anisotropic sectors one could propose. As for the issues
of scaling behavior, due to the  limitation of small Reynolds numbers
in the numerical case (\cite{ara99}),
 and to the limited amount of information available
on the tensorial structure of the velocity field in the experimental case 
(\cite{ara98,kur00}) only partial answers have been found.  Among them,
the most important is the strong universality shown by the isotropic sector
as a function of the local degree of non-homogeneity (and anisotropy),
i.e. the strong universality showed by the scaling
 properties of the coefficients
$a_{q,00}(\x,R)$ as a function of $\x$ in non-homogeneous 
turbulence (\cite{ara99}).
On the other hand, in this paper we would like to also 
propagate the SO(3) decomposition as an appropriate tool to analyze,
characterize, and quantify the non-universal large scale 
geometric properties of the turbulent flow.\\
 As an example we take 
numerical channel flow (\cite*{ama97,tos99}) obtained by a lattice
Boltzmann code running on a massively parallel machine. The spatial
resolution of the simulation is $256\times 128 \times 128$ grid 
points. Periodic boundary conditions were imposed along the 
stream-wise (x) and span-wise (z) directions, whereas no slip boundary
conditions were applied at the top and at the bottom planes (y-direction). 
The Reynolds number at the center of the channel is about 3000. 
 In our case
of channel flow 
we assume that, due to the homogeneity in planes parallel to the walls, 
 there is only a dependence on the height $y$ of all statistical observable. 
The coefficients $a_{q,jm} (\x, R)$
carry two types of information: (i) Their scaling behavior 
$a_{q,jm} (\x , R)\propto R^{\zeta^{(2)}_{q,jm}}$
which at least for small scales and large $Re$ is hoped to be universal,
i.e., position and flow independent\footnote{The issue of 
universality of sectors with $j>0$ is far from being trivial. A lack
of universality may be due to the existence of infrared (IR) or ultraviolet (UV) divergences in the non-local integral induced by the
 pressure terms in the Navier-Stokes equations, \cite{ara98}.} 
and (ii) their absolute or relative magnitudes which clearly are non-universal, i.e., position 
$\x$ and flow type dependent. These ratios characterize 
what kind of structures the flow contains. These are time and 
ensemble averaged quantities, obeying the underlying Navier-Stokes
SO(3) symmetry, and we consider them to be a more systematic 
tool for structures characterization than snapshots of vortex sheets,
worms, swirls or contour plots of either the velocity or the 
vorticity fields. 

When analyzing higher order structure tensors  $C_{\alpha \beta \dots \gamma}
(\x , \R)$ the decomposition of type (\ref{eq2}) becomes cumbersome soon.
Moreover, in most experiments the full tensorial information is not
available anyhow. Therefore,
one has to restrict oneself to an abbreviated form of the SO(3) decomposition
of the velocity structure tensor, namely, the SO(3) decomposition 
of the longitudinal structure function. 
In this case, being the undecomposed observable
a scalar under rotations, there exists only one irreducible representation
for any $j$ sector, i.e. 
the usual spherical harmonics basis set $Y_{jm}(\unitR)$. We decompose the
longitudinal structure function 
\be
S_L^{(p)}(\x,\R)=
\la\left(\lp\u\lp\x+\R\rp-\u\lp\x-\R\rp\rp\cdot\unitR\right)^p\ra
\label{eq4}
\ee
as follows:
\be
S_L^{(p)}(\x,\R)=\sum_{jm}{S_{jm}^{(p)}(\x,R)Y^{jm}(\unitR)}.
\label{eq5}
\ee
We expect that when scaling behavior sets in (presumably at high enough Re)
we should find:
\begin{equation}
 S^{(p)}_{jm}(\x,R) \sim a_{jm}(\x) R^{\zeta_{jm}^{(p)}} \ .
\label{scale}
\end{equation}
Again, the  $a_{jm}^{(p)} (\x , R)$
carry both the  scaling information 
 $a_{jm}^{(p)} (\x , R) \propto R^{\zeta_{jm}^{(p)}}$
and their non-universal amplitudes.\\
A practical problem with the
decomposition (\ref{eq5}) of (\ref{eq4}) is 
that for $\x$ close to the boundaries the scale $R$ is 
restricted to lengths smaller than the distance from the wall\footnote{Because
of the trivial remark that the analyzing sphere cannot touch the walls.}.
More generally,
 $R$ cannot exceed a 
typical distance over which non homogeneities are overwhelming.
Therefore we will also perform a decomposition of (\ref{eq4})
which obeys the weaker SO(2) symmetry, i.e. rotational invariance
in a plane for fixed distance $y$ from the wall,
\be
D_L^{(p)} (y, \R ) = \sum_{m} d_{m}^{(p)} (y , R) \exp{(im\phi )}.
\label{eq6}
\ee
The orientation dependence in a plane reduces to the dependence on an
angle $\phi$. Again, the 
$d_{m}^{(p)} (y,R)$ carry both scaling and amplitude information.

Let us notice at this point, that  the SO(3) decomposition has its roots
on the intimate structure of the Navier-Stokes eqs, i.e on the invariance
under rotations of the inertial and dissipative terms and on the relative
foliations on different sectors of the rotational group
of the equation of motion of any correlation function \cite{ara99b}.
 On the other hand,
do not exist closed equations for two-dimensional observable and
therefore  the SO(2) decomposition can  only be
 seen as a  powerful tool to exactly decompose any observable in a fixed
plane as a function of isotropic and anisotropic structures in 
the plane itself.
Clearly, such a kind of decomposition can teach us a lot in those regions, like
at the border between  the viscous and turbulent  boundary layers, where
strongly anisotropic but planar structures named 'streaks' are 
supposed to carry the most important dynamical information of the flow.  

\section{SO(3) analysis of a turbulent  channel flow field}
In previous studies most of the attention was paid on the isotropic
sector of the structure function decomposition (\ref{eq5}),
 i.e. on the behavior of $ S_{00}^{(p)} (\x , R)$
as a function of the center of the decomposition, $\x$, and of the
scale $R$. In \cite{ara99} it was showed that the 
isotropic projection enjoys much better scaling properties
than the undecomposed structure function and that these properties 
are robust  with respect to the changing of the 
 local degree of anisotropy, i.e. with
respect to the center of the decomposition, $\x$. 
These findings support the idea of universality of the 
isotropic scaling exponents.Very little
was possible to say about scaling of the anisotropic sectors because of
lack of spatial resolution; the only qualitative statement was that
the scaling exponent of the $j=2$ sector was roughly $4/3$, as predicted by
the dimensional argument given by \cite{lum67} or by \cite{gro94}.\\
Here we want to concentrate also on the more applied question of how much 
the different projections, independently on their possible
scaling properties, can teach us about the preferred geometrical 
structures present in the flow at changing the 
analyzing position in the channel. \\
In figs. \ref{fig1}  and \ref{fig2} we present the three different contributions
 we have in the $j=2$ non-isotropic sector\footnote{The $j=1$ sector
is absent due to the symmetries of the structure functions chosen
in this work.}
extracted at the center of the channel ($y^+=160$) and at one
 quarter ($y^+=80$)
respectively.\\
The relative size of the $S_{2m}^{(2)} (y^+,R^+)$ for different $m$ and 
fixed
$y^+$ characterize the geometry of the anisotropic structures on the
corresponding scale $R$. E.g., for $y^+=80$ the $(j=2,m=1)$  mode
 is
very pronounced on smaller scales, see fig.\ \ref{fig2}. We associate
this with the hairpin vortices and other structures which diagonally
detach from the wall and which are projected out by $Y_{21}$. For
a visualization of the $Y_{2m}$ see figure \ref{fig3}. 
In the center the $(j=2,m=1)$ mode is two orders of magnitude
less pronounced than at
$y^+=80$. Our interpretation is that the diagonal structures from above
and below have equal and opposite contributions.\\
The most pronounced structures in the center are those parallel to 
the flow direction, i.e., $(j=2,m=2)$, see figure \ref{fig1}. \\
Also at $y^+=80$ the structures parallel to the flow direction 
(mode $(j=2,m=2)$) are rather pronounced. At scales beyond $R^+ \approx 100$
they overwhelm the diagonal contributions (mode $(j=2,m=1)$). Therefore
one is tempted to interpret $R^+ \approx 100$ as the maximal size (in average)
of the hairpin vortices.\\
We re-did this type of analysis also for the $S_{2m}^{(4)} (y^+,R^+)$ with
very similar results.
\subsection{Higher order moments and the lack of isotropy at small scales}
The first question one may want to ask
 about the decomposition (\ref{eq5})
is whether it {\it converges} with increasing $j$. We want to check
this for an $\R$ in (stream-wise) flow direction, i.e., 
$\unitR = (\theta , \phi )=(\pi/2 , 0)$. 
As we can see from fig.\ \ref{fig4}, at small scales
and in the channel center, where anisotropic contributions 
are small, the convergence
is rather good. But away from the center ($y^+=62$) and in 
particular for large scales  quality of the convergence 
become poor, see fig. \ref{fig5}. 
Note that in any case the convergence is not
monotonous as a function of the scale.
This is a systematic quantitative way to understand the rate of isotropization
toward small scales exhibited by this particular flow as a function
of the distance from the wall. 

Another, even more informative way to quantify the rate of isotropization is 
to plot the ratio of each single amplitude $S_{jm}^{(2)}(\x,R)$ 
to the total
structure function $S_L^{(2)}(\x,\R)$ with $\R$ in the direction of the 
mean flow. In figs. \ref{fig6} and \ref{fig7} 
one can find the above quantities at the center
of the channel  
$y^+=160$ and in the buffer layer $y^+=62$, respectively. 
 What is very interesting
to notice is that at large scales there are
 contributions from all resolved $j$ sectors
indicating as expected a lack of convergence of the
 decomposition at those scales
and that in the buffer layer the relative ratio of the
 anisotropic sectors is much higher than 
what is seen in the center. Moreover, even more interesting,
 in the buffer layer, where
due to the presence of a high shear one can imagine a statistically 
stable signature
of anisotropic physics there appears a clear grouping of 
different sectors labeled by
different $j$ indexes. Fig. \ref{fig7} shows that  projections
with the same $j$ but different $m$ 
indexes have a qualitative similar behavior.
Of course, these kind of comparison depends on the direction of the 
undecomposed  structure functions (here taken parallel to the walls). \\
Another possible test of the  relative weights of anisotropies, 
free of the previous arbitrariness, 
 is to plot the ratio between the isotropic projection 
$S_{00}^{(p)}(\x,R)$
 and  the other anisotropic projections for $j>0$. Such a test is done
in terms of quantities depending only on the separation magnitude $|R|$,
and therefore measures the relative importance of anisotropies independently
of the orientation. 
In Fig \ref{fig8} we show, for example,
the ratio between the sector $(j,m)=(4,4)$ and the isotropic 
sector $(j,m)=(0,0)$
at changing the analyzed height in the channel and for 
all $R^+$. As it is possible
to see, as expected, by approaching the wall (decreasing $y^+$) 
the ratio becomes
larger and larger, showing clearly the importance of high $j$ fluctuations
in the sheared buffer layer. \\
All the previous trends  have also been found, amplified, by analyzing
higher moments. For example, in figs. \ref{fig9} and \ref{fig10} we 
re-plot the same of figs. \ref{fig6} and \ref{fig7}
 but for the fourth order structure functions. 
The fact that the previous trends
are much more enhanced for higher order moments is a clear 
indication that anisotropy fluctuations are important but 'rare', i.e.
are connected to persistent  intense fluctuations in a sea
of isotropic turbulence. 

\section{SO(2) analysis of a turbulent channel flow}
As extensively discussed
in the previous sections, the SO(3) decomposition turned out
to be extremely useful from both its theoretical background 
connected to the symmetry of the NS eqs and its ability 
to highlights statistical information as a function of their
geometrical structures. On the other hand, the SO(3) decomposition
suffers from some drawbacks when one wants to analyze the statistical
turbulent behavior close to the fluid boundaries. This is
due to the obvious fact that in order to perform the decomposition
one needs to perform integrals over a given sphere, and therefore
close to the boundaries the limitation of the sphere radius 
does not allow to extract any information but for a very
limited (almost fully dissipative) range of scales. \\
To overcome this problem we propose to use
a decomposition in eigenfunction of the group of rotations
in two dimensions, SO(2). 
The rational behind this idea is that the Navier Stokes equations
obviously obey the $SO(2)$ symmetry and for the channel flow also the geometry
obeys this symmetry, once the rotation axis is chosen in the $y$ direction.
However, the mean flow breaks the $SO(2)$ symmetry as it breaks the 
$SO(3)$ symmetry. 
Nevertheless we will gain a tool being able to exactly decompose any 
two-dimensional observable in terms of fluctuations with a given property under
two-dimensional rotations. 
In the region very close to the walls
where very elongated 'streak' structures have been observed,
the SO(2) analysis may help in understanding the relative
importance of isotropic (in the plane) and anisotropic (in the plane)
fluctuations. \\
Another very important issue we want to address by using the SO(2)
decomposition is connected to the recent findings by  \cite{tos99}
of a different
intermittent behavior close to the walls ($y^+\sim 35$) 
shown by  longitudinal structure functions in the stream-wise direction. 
These results were also connected to  the breaking of the Refined Kolmogorov Similarity Hypothesis (RKSH) in the buffer layer, \cite{fede_pof}. 
Here, we show that by means
of the SO(2) decomposition we are able to highlight the importance of
the streak like structures in determining this higher intermittent
behavior.\\
In this section 
we are interested only in observable in planes
parallel to the walls and therefore the SO(2) decomposition of, say,
the longitudinal structure function is defined as 
\be
D_L^{(p)} \lp y , \R \rp = \sum_{m} d_{m}^{(p)} \lp y, R\rp \exp{(\lp im\phi \rp)},
\label{eq:2d}
\ee
where $\R$ is a two-dimensional vector lying in a plane at fixed $y$,
and $D_L^{(p)} (y , \R )$ is the longitudinal structure function
in the direction $\R$. Due to the symmetry of the structure function
we expect that only even $m$s will contribute to the sum in (\ref{eq:2d})
\\
In Fig. \ref{fig11} and Fig. \ref{fig12}
we show the rate of convergence of the reconstructed structure function
of order 2  as function of the maximum $M$ contributing to
the right hand side (RHS) of (\ref{eq:2d}) and at two different distances 
from the wall, at the center (fig. \ref{fig11}) and in the buffer layer 
(fig. \ref{fig12}). As it is clear,
again we find a quite good convergence in the center. In the 
buffer layer, especially large scales are still far from being reconstructed
even reaching $M=8$. 
This is a clear tendency of formation of very
large and intense structures in the strongly anisotropic buffer. 
These trends are
even more pronounced for the fourth order moment as shown in Figs. \ref{fig13} 
and \ref{fig14}.\\
In Fig. \ref{fig15}  we show the absolute weight 
of different $m$-contributions for the second order structure function
again in the center. It is interesting to notice
that there is a clear monotonic organization of different
contributions as a function of their isotropic/anisotropic properties, i.e.
higher $m$s are less intense than lower $m$s systematically way at all scales.\\
On the other hand, in the buffer layer, Fig. \ref{fig16}, 
there is a crossing of the
$m=2$ contribution and of the $m=4$ contribution at scales
 of the order of $R^+ \sim 90$.
We interpret this crossing as the evidence of the
 formation of structures with typical
size $R^+ \sim 90$ and with a preferred 
orientation given by the $m=2$ eigenfunction.
The $m=2$ eigenfunction weights preferentially structures with a positive correlation
in the stream-wise direction and negative correlated in the span-wise direction, i.e.
exactly 'streak' like structures as those observed also in our numerical simulation
by performing simple contour plots (see Fig. \ref{fig17}).
 As it is always the case, the above
observed trends are even more intense for 
$p=4,6..$. For $p=6$ it even happens (not shown) that the dominant
contribution at large scales 
is given by the $m=2$ sector, proving, once more, the extreme departure 
from isotropy (in the plane) close to the walls.\\ In order to quantify the 
departure from isotropy in each planes at changing the distance from the wall
we plot  the ratios between the
projections on the $m=2$ sector and the isotropic sector (Fig. \ref{fig17a})
 at varying the distance
from the wall and for some $R^+$ values.
 Fig.\ \ref{fig17b} shows the same but for $m=4$.
 It is interesting to notice, how there is a sharp
transition for $y^+ \sim 40$ from an almost isotropic statistics ($y^+ > 40$)
and a strongly anisotropic statistics ($y^+ < 40$), again
the clear signature of the beginning of a ``structured'' buffer for
$(y^+ < 40)$ shows up.

\subsection{Review of near wall physics}
Let us now switch to the  more statistically
 minded question connected to the existence 
of different intermittent properties close to the walls as previously 
reported in \cite{tos99}.
This issue is connected to the general question whether in strong shear regions
for a range of scales larger than the typical 
shear length, $L_{\S}=\lp\varepsilon/\S^3\rp^{1/2}$,
 one may have a different
statistic transfer of energy than what expected at scales smaller than the typical
shear length. Different statistical energy transfer
 properties would have as a consequence
also different intermittent properties of the velocity structure functions and, probably,
the breaking of the RKSH if the shear lengths is small enough to be comparable
with the dissipative length. Only in very high sheared region one can 
hope to have  a very small $L_S$ and therefore a sufficient range of scales with $R >L_S$ 
where scaling laws can be investigated. Indeed in many different sheared 
flows, 
\cite{oci00,fed2,fed3}, it has been found that in strongly sheared region intermittency increase and display universality (i.e. same exponents were measured in very different set-up).
In \cite{tlc00} it was given a simple theoretical explanation of how intermittency should change in presence of strong shear.
Considering the usual decomposition of the velocity field in its average and fluctuating part, $\bv(\bx;t)={\overline {\bv}}(\bx)+\bv'(\bx;t)$, we get (from the Navier-Stokes equations) the usual Reynolds decomposition
\be
D_t v'_i+\S_{ij}(\bx) v'_j +
 v_j' \partial_j v'_i - {\overline {v_j' \partial_j v'_i}}
= - \partial_i p' + \nu \Delta v'_i
\label{reynolds}
\ee
with $D_t=\lp \partial_t + {\overline v}_j \partial_j \rp$.
The shear is defined as
$\S_{ij}\lp\bx\rp= \partial_j {\overline v_i\lp\bx\rp}$
and depends on the mean flow geometry.

In equation \ref{reynolds} the second and third terms of left hand side (LHS) will be of the same order at a scale $L_{\S} \sim \lp \epsilon / S^3 \rp^{1/2}$ (shear length scale).
For scale smaller than $L_{\S}$ it is the third term in eqn. \ref{reynolds} that will balance the energy dissipation, and hence the usual Refined Kolmogorov Similarity Hypothesis will hold: $S_p(r) \sim \la
\varepsilon(r)^{p/3}\ra \cdot r^{p/3}$. For scales larger than $L_{\S}$ it will be the second term in eqn. \ref{reynolds} that will balance the energy dissipation and hence: $S_p(r) \sim \la \varepsilon(r)^{p/2}\ra$. The validity of this second relation i
n region of high shear values was already established by
 \cite{fede_pof}.\\
We want now to see how much one can say about this new 'intermittent' behavior
close to the channel walls. In order to extract any quantitative 
information on scaling exponents in numerical simulation one needs to 
use the ESS
technique, \cite{ben93b,ben96b,gro97b}. ESS is based on the experimental and numerical observation
that structure functions even at moderate Reynolds numbers show scaling in a generalized
sense, i.e. by studying the relative scaling of one structure functions, say the
second order structure function, versus any other. In particular, we want to 
verify and exploit that the following scaling holds:
\begin{equation}
D_L^{(p)} (y , \R ) \sim  \lp D_L^{(2)} (y , \R )\rp^{\zeta_y^{(p)}
/\zeta_y^{(2)}}
\label{eq:ess}
\end{equation}
where we have again limited ourself to the analysis of structure functions in the plane.
In (\ref{eq:ess}) we have explicitly taken into account the possibility that the scaling
exponents depend on the distance from the walls. In particular,  \cite{tos99}
 showed, by analyzing the same data set, that there exist two distinguished
set of exponents. One governing the scaling in the range of scales smaller than $L_S$
(i.e. close to the center of the channel, in our case) which is given in terms
of the usual isotropic
and homogeneous set of exponents. The second governing the scaling
in the sheared range of scales $R >L_S$ (i.e. close to the walls in our channel simulation)
which is given in terms of a much more intermittent set of exponents. \\
In Fig. \ref{fig18} 
we show the ESS local slopes of the undecomposed structure function
in the stream-wise direction in the center of the channel and 
the ESS local slope
of the projection on the $m=0$ sector always at the center 
of the channel for the moments
$p=2$ versus $p=4$. As it is evident, already the fully isotropic 
component (in the plane)
is able to well reproduce the undecomposed observable and are both in good agreement
with the isotropic and homogeneous scaling. Of course, the previous finding
confirms the simple statement that at the center of the channel the shear length is formally infinite and therefore the whole range of scales 
available is weakly affected by
any shear effect. On the other hand in Fig. \ref{fig19} we show 
the same quantities
of Fig. \ref{fig18}
 but in a plane well inside the buffer layer ($y^+=37$). As it is possible
to see now the $m=0$ component does not reproduce
 the undecomposed observable, confirming
the evident fact that here we are strongly anisotropic. 
Nevertheless, it is enough
to add the $m=2$ sector, i.e. to 
reconstruct up to $M=2$ in the RHS of (\ref{eq:2d})
to have a very good agreement with the more intermittent 
undecomposed structure functions local slope. This is a good quantitative evidence that
as far as the new  scaling properties  are concerned the main effects is brought by
these $m=2$ 'streak' like structures in the buffer layer. \\
Still we need to understand the physics of these structures, why they are
 more intermittent,
whether it is a coincidence or not that the new set of exponents coincides
extremely well with what measured for passive scalar
 advected by a turbulent flow \cite{cha95}. Nevertheless,
 we are confident that having a systematic way
to analyze any isotropic/anisotropic 
two-dimensional/three-dimensional turbulent data
set may help in further advance of the field.

\section{Conclusions}
A detailed  investigation of anisotropies in channel flows
in terms of the SO(3) and SO(2) decomposition 
of structure functions has been presented. 
Projections on the eigenfunction of the two symmetry groups
can be seen as a systematic 
expansions of structures as a functions of their scale 
and in terms of their local degree
of anisotropy. \\ We have used the SO(3) decomposition of structure functions
at the center and at one quarter of the channel in order to have
a quantitative tool to measure the relative importance
of isotropic and anisotropic fluctuations at all scales. Close to the
wall, the anisotropic fluctuations show strong effects induced
from structure with the typical orientations of {\it hairpin} vortices.
A
partial lack of isotropization is still detected at the smallest resolved
scales. \\
The SO(2) decomposition in planes parallel to the walls allowed us
to access also the viscous and buffer regions. In those regions,
we have found that the strong enhancement of intermittency 
can be understood in terms of {\it streak} like structures
and their signatures in some coefficients of the SO(2) decomposition. \\
We think that the method  presented here 
is beneficially applicable in all those
cases where quantitative
comparison and/or  studies of anisotropic effects in different flows are
needed (channel flows, boundary layers, homogeneous shear etc..) \\ 
The application of similar decomposition  to small scales
observable like  vorticity and  energy dissipation would certainly
be of great interest too. \\ 
Having the possibility to control the anisotropic behavior
is of great importance to improve LES of strongly anisotropic 
and inhomogeneous
flows.

\vspace{1cm}

\noindent
{\bf Acknowledgments:}
The authors thank I. Arad for I. Procaccia for helpful discussions. 
The work is part of the research  program of the Stichting voor 
Fundamenteel Onderzoek der Materie (FOM), which is financially supported 
by the Nederlandse  Organisatie voor Wetenschappelijk Onderzoek (NWO).
This research was also supported in part
by the European Union under contract HPRN-CT-2000-00162 and
by the National Science
Foundation under Grant No. PHY94-07194 and we thank the Institute
of Theoretical Physics in Santa Barbara for its hospitality.

\section{Appendix A}
In this appendix, we want to explicitly write down the 
SO(3) decomposition of the most general two-point velocity 
correlations in anisotropic  turbulence. 
We consider the second order tensor
involving velocities at two distinct points ${\bf x_1} = {\bf x}$
and ${\bf x_2} = {\bf x+R}$:
\begin{equation}
C^{\alpha \beta }(\R) \equiv \left\langle u^{\alpha }(\R)u^{\beta }(\R+
\x)\right\rangle \;.
\label{tensor}
\end{equation}
where we have supposed that the statistics is homogeneous (but not isotropic)
and therefore the LHS  of (\ref{tensor}) depends only
 on ${\bf R}$, the distance
between the two points. 
Then, we can decompose $C^{\alpha \beta}$ according
to the irreducible representations of the SO(3) groups.
Each irreducible representations
will be composed  by a set of functions labeled with the usual 
 indices $j=0,1,....$ and $m= -j, ..., +j$ corresponding to the total 
angular momentum and to the projection of the total angular momentum on
a arbitrary direction respectively. Moreover, 
 a new 'quantum' index $q$ which  labels
 different irreducible representations will be necessary.  It is easy 
to realize that  there are only $q=1,...,9$  irreducible
representations of the SO(3) groups on the space of two-indices
tensor depending continuously from a three-dimensional vector, \cite{ara99b}.
In particular, fixed $j$ and $m$, 
 the $9$ basis tensor can be simply constructed
starting from the scalar spherical harmonics $Y_{j,m}(\hat{\bf X})$
plus successive application of the two isotropic operators
$R_{\alpha}$ and $\partial_{\beta}$ in order to saturate the correct
number of tensorial indices. For example, the $9$ linearly
independent basis vectors which defines the irreducible representations
in our case can be chosen as:

\begin{eqnarray*}
B_{1,jm}^{\alpha \beta }(\unitR)&\equiv& R^{-j}\delta
^{\alpha \beta }\Phi_{jm}(\R)\;, \\
B_{2,jm}^{\alpha \beta }(\unitR)&\equiv&
R^{-j+1}\epsilon ^{\alpha \beta \mu }\partial _{\mu }\Phi_{jm}(\R)\;,\\
B_{3,jm}^{\alpha \beta }(\unitR)&\equiv&
R^{-j}[x^{\alpha} \partial^{\beta} -R^{\beta} \partial^{\alpha}]
\Phi_{jm}(\R)\;,\\
B_{4,jm}^{\alpha \beta }(\unitR)&\equiv&
R^{-j-1}\epsilon ^{\alpha \beta \mu }R_{\mu }\Phi_{jm}(\R)\;, \\
B_{5,jm}^{\alpha \beta }(\unitR)&\equiv&
R^{-j+2}\partial ^{\alpha }\partial^{\beta }\Phi_{jm}(\R)\;,\\ 
B_{6,jm}(\unitR)&\equiv& R^{-j+1}[\epsilon ^{\beta \mu
\nu }R_{\mu }\partial _{\nu }\partial ^{\alpha }+\epsilon ^{\alpha \mu \nu
}R_{\mu }\partial _{\nu }\partial ^{\beta }]\Phi_{jm}(\R)\;, \\
B_{7,jm}^{\alpha \beta }(\unitR)&\equiv&
R^{-j}(x^{\alpha }\partial ^{\beta }+R^{\beta }\partial ^{\alpha })
\Phi_{jm}(\R)\;, \\
B_{8,jm}^{\alpha \beta }(\unitR)&\equiv&
R^{-j-1}[x^{\alpha }\epsilon ^{\beta \mu \nu }R_{\mu }\partial _{\nu
}+R^{\beta }\epsilon ^{\alpha \mu \nu }R_{\mu }\partial _{\nu }]\Phi_{jm}(\R)\;, \\
B_{9,jm}^{\alpha \beta }(\unitR)&\equiv&
R^{-j-2}x^{\alpha }R^{\beta }\Phi_{jm}(\R)\;. \\
\end{eqnarray*}
Where for the sake of simplicity we have posed 
$\Phi_{jm}(\R) \equiv  R^j \,Y_{jm}(\unitR)$.
As a results the most general second order tensor like (\ref{tensor}) 
can be decomposed as:
\begin{equation}
C^{\alpha \beta }(\R) \equiv 
\sum_{j,m} \sum_{q=1}^9 c_{q,jm}(R) B^{\alpha \beta}_{q,jm}(\unitR)
\end{equation}
where now the physics of the anisotropic statistical fluctuations
must be analyzed in terms of the projections $c_{j,m,q}(R)$
in the different sectors.

\bibliographystyle{jfm}
\bibliography{blmt.bib}

\newpage

\begin{figure}
\epsfxsize=.8\hsize
{\hskip 0.0cm{\centerline{\epsfbox{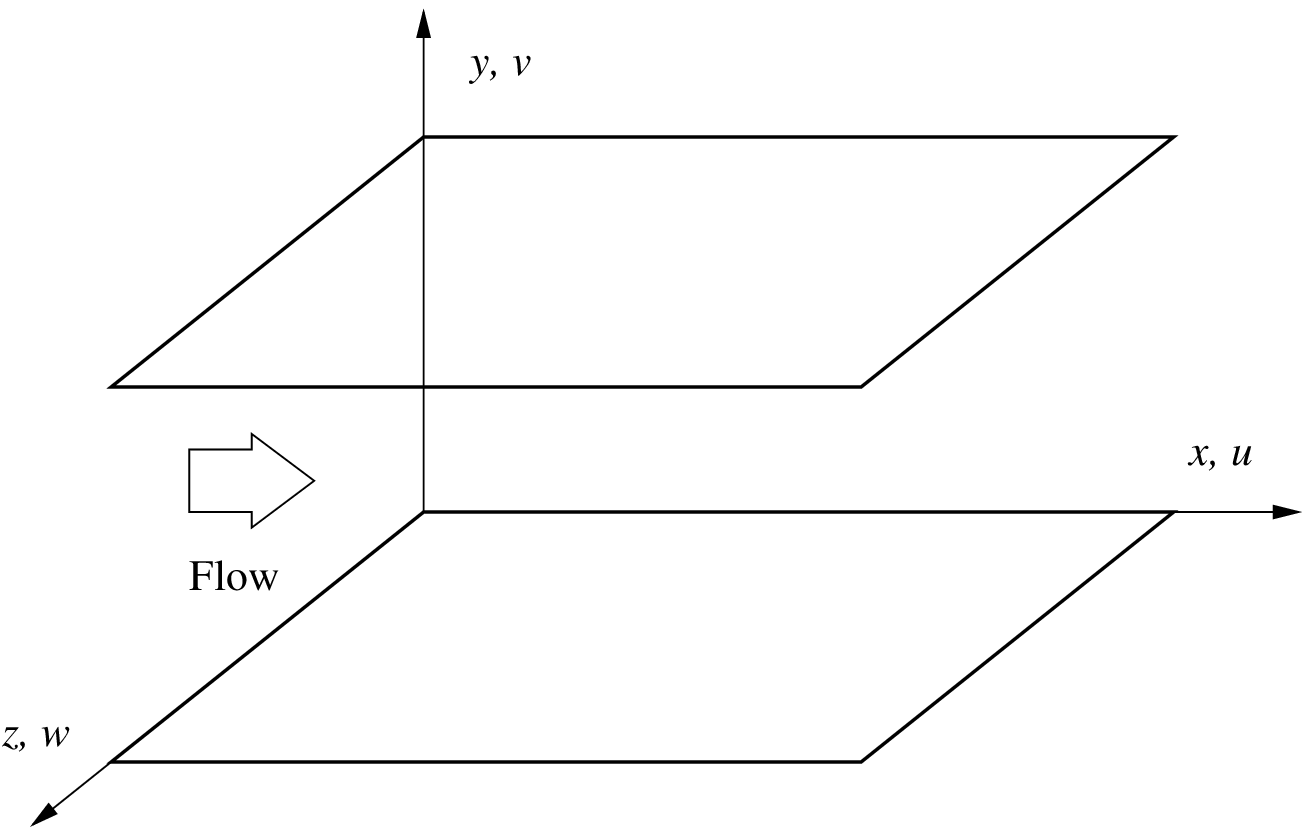}}}}
\caption{Coordinate system in channel. Shown are stream-wise ($x$), span-wise ($z$) and wall-normal ($y$) directions.}
\label{fig0}
\end{figure}

\begin{figure}
\epsfxsize=1.1\hsize
{\hskip -.8cm{\centerline{\epsfbox{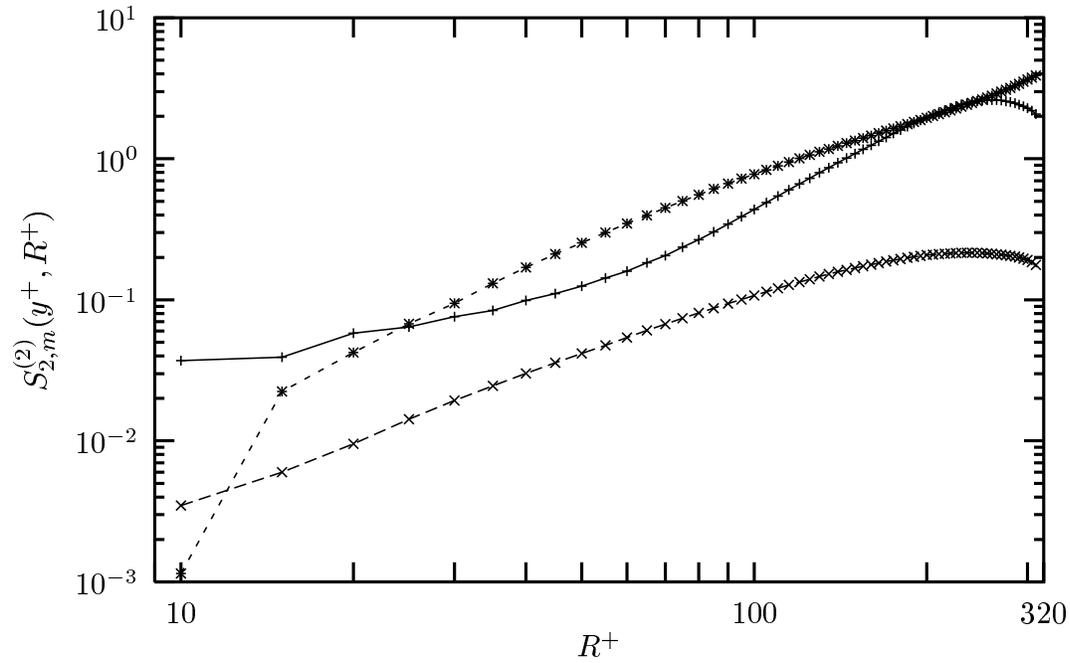}}}}
\caption{Log-log plot of $S^{(2)}_{2,2}(y^+,R^+)~(\ast)$; $S^{(2)}_{2,1}(y^+,R^+)~(\times)$ and $S^{(2)}_{2,0}(y^+,R^+)~(+)$ as functions of $R^+$ at the center of the channel $y^+=160$.}
\label{fig1}
\end{figure}
\begin{figure}
\epsfxsize=1.1\hsize
{\hskip -0.8cm{\epsfbox{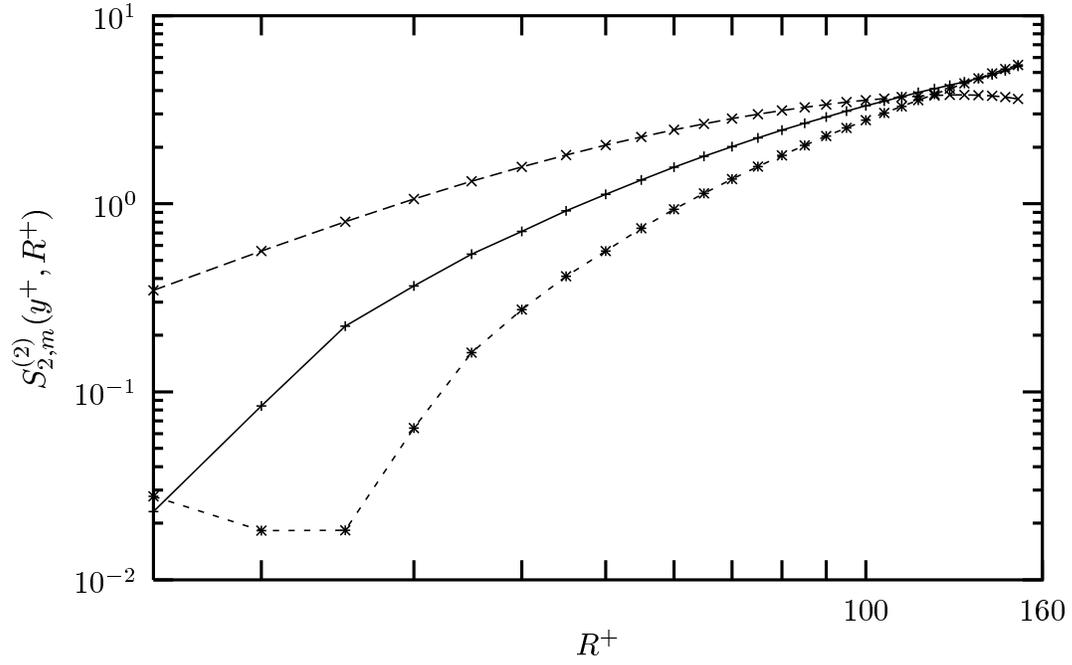}}}
\caption{Log-log plot of $S^{(2)}_{2,2}(y^+,R^+)~(\ast)$; $S^{(2)}_{2,1}(y^+,R^+)~(\times)$ and $S^{(2)}_{2,0}(y^+,R^+)~(+)$ as functions of $R^+$ at $y^+=80$.}
\label{fig2}
\end{figure}

\begin{figure}
\epsfxsize=1.1\hsize
{\centerline{\epsfig{file=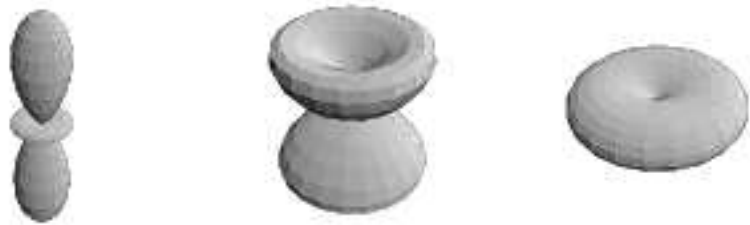}}}
\caption{Graphical representation of spherical harmonics (from left to right) $\left|Y^{2,0}(\theta,\phi)\right|$, $\left|Y^{2,1}(\theta,\phi)\right|$ and $\left|Y^{2,2}(\theta,\phi)\right|$. }
\label{fig3}
\end{figure}

\begin{figure}
{\epsfxsize=1.05\hsize{\hskip -0.2cm{\epsfbox{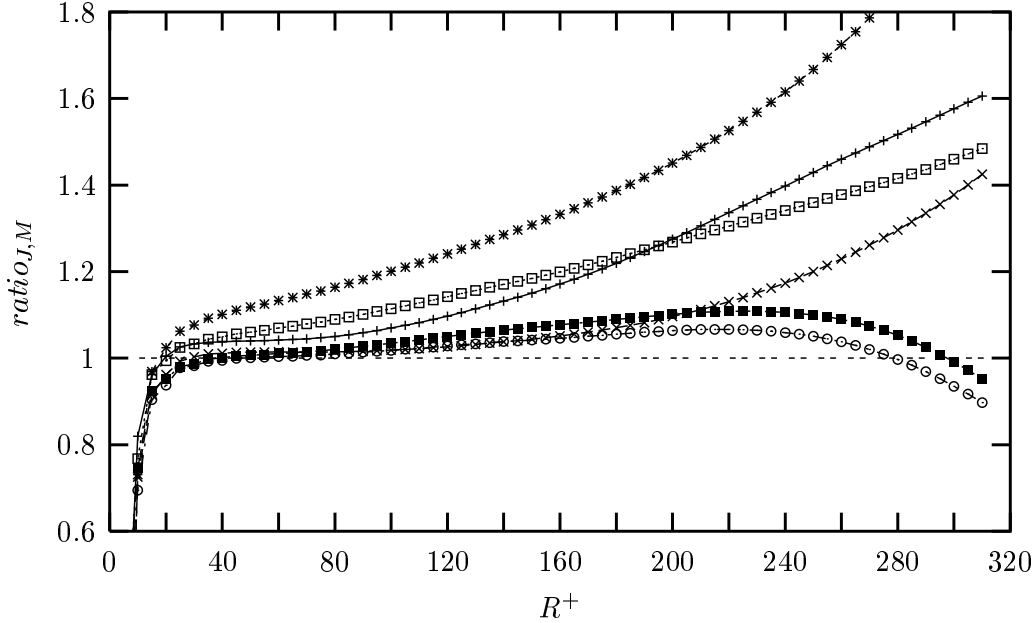}}}}
\caption{\label{fig4}Analysis of the convergence of the $SO(3)$ decomposition:
$ratio_{J,M}$ represents the ratio between the longitudinal structure function of order $2$ in the stream-wise direction reconstructed up to $(J,M)=(0,0)~(+)$; $(J,M)=(2,0)~(\times)$; $(J,M)=(2,2)~(\ast)$; $(J,M)=(4,0)~(\square)$; $(J,M)=(4,2)~(\blacksquare)
$ and $(J,M)=(4,4)~(\circ)$ and the undecomposed structure function, at the center of the channel $y^+=160$.} 
\end{figure}

\begin{figure}
\epsfxsize=1.05\hsize{\hskip -0.2cm{\epsfbox{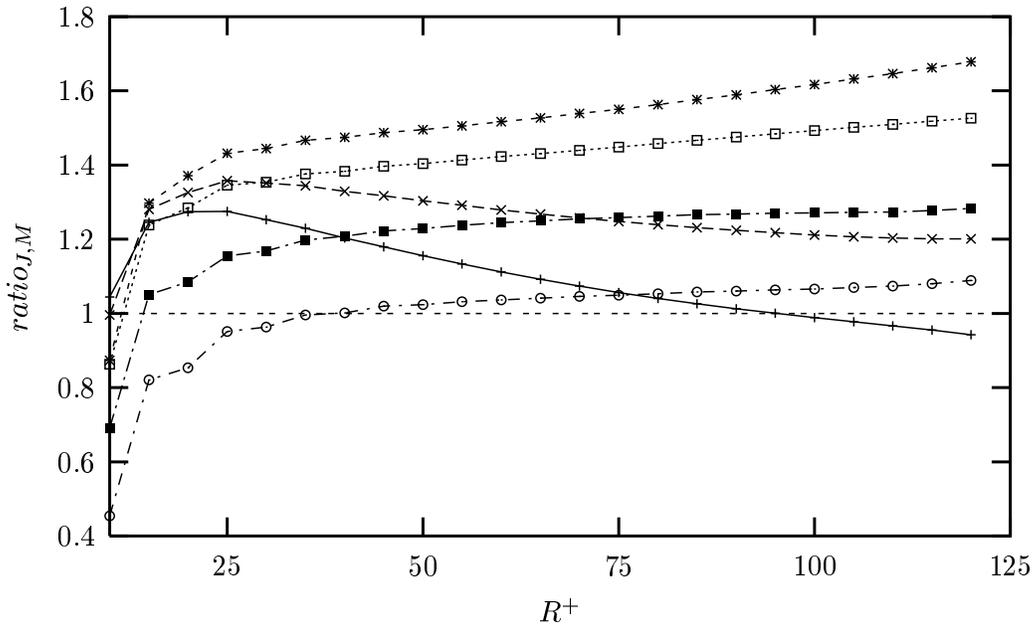}}}
\caption{Analysis of the $SO(3)$ decomposition's convergence:
$ratio_{J,M}$ represents the ratio between the longitudinal structure function of order $2$ in the stream-wise direction 
reconstructed up to $(J,M)=(0,0)~(+)$; $(J,M)=(2,0)~(\times)$; $(J,M)=(2,2)~(\ast)$; $(J,M)=(4,0)~(\square)$; $(J,M)=(4,2)~(\blacksquare)
$ and $(J,M)=(4,4)~(\circ)$ and the undecomposed structure function, at the center of the channel $y^+=62$} 
\label{fig5}
\end{figure}

\begin{figure}
\epsfxsize=1.2\hsize
{\hskip -.8cm{\centerline{{\epsfbox{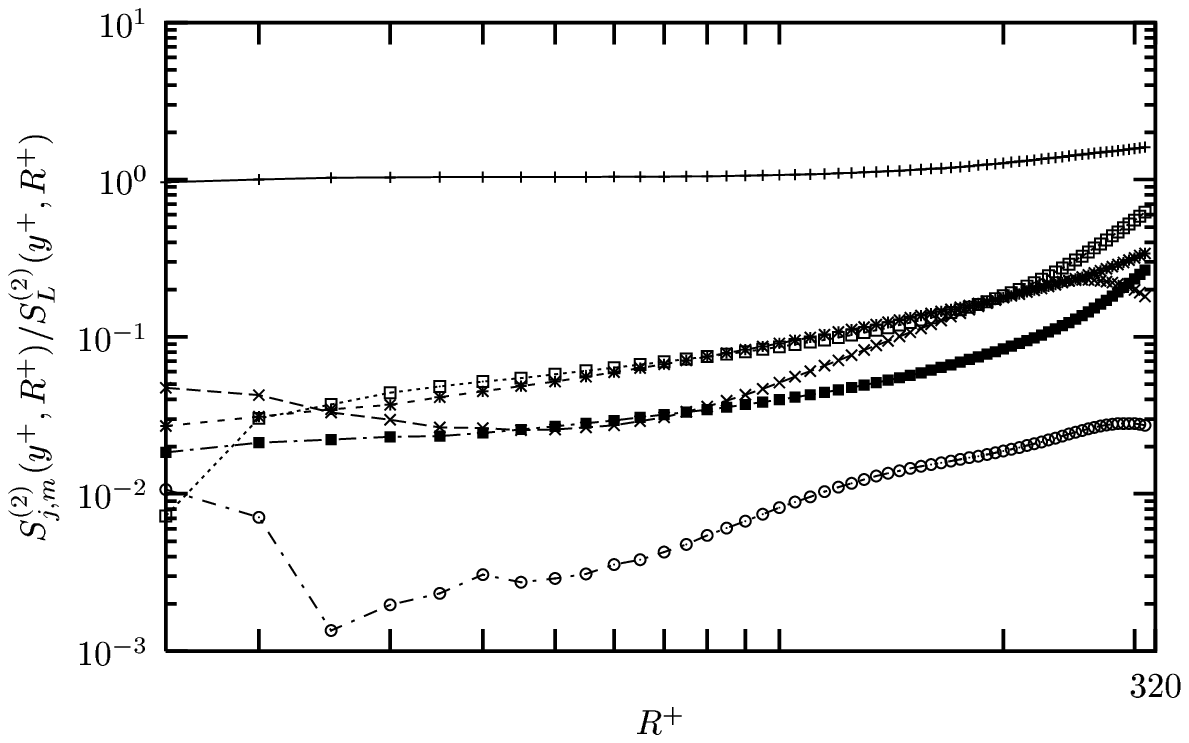}}}}}
\caption{Ratio of each single $(j,m)$ amplitude $S^{(2)}_{j,m}(y^+,\R)$ 
to the total structure function $S^{(2)}_L(y^+, \R)$ with ${\bf R}$ in 
the direction of the mean flow and $y^+=160$. The $(j,m)$ indexes are: 
$(0,0)~(+)$; 
$(2,0)~(\times)$; $(2,2)~(\ast)$; $(4,0)~(\square)$; $(4,2)~(\blacksquare)$ and $(4,4)~(\circ)$.}
\label{fig6}
\end{figure}

\begin{figure}
\epsfxsize=1.2 \hsize
{\hskip -0.8cm{\centerline{{\epsfbox{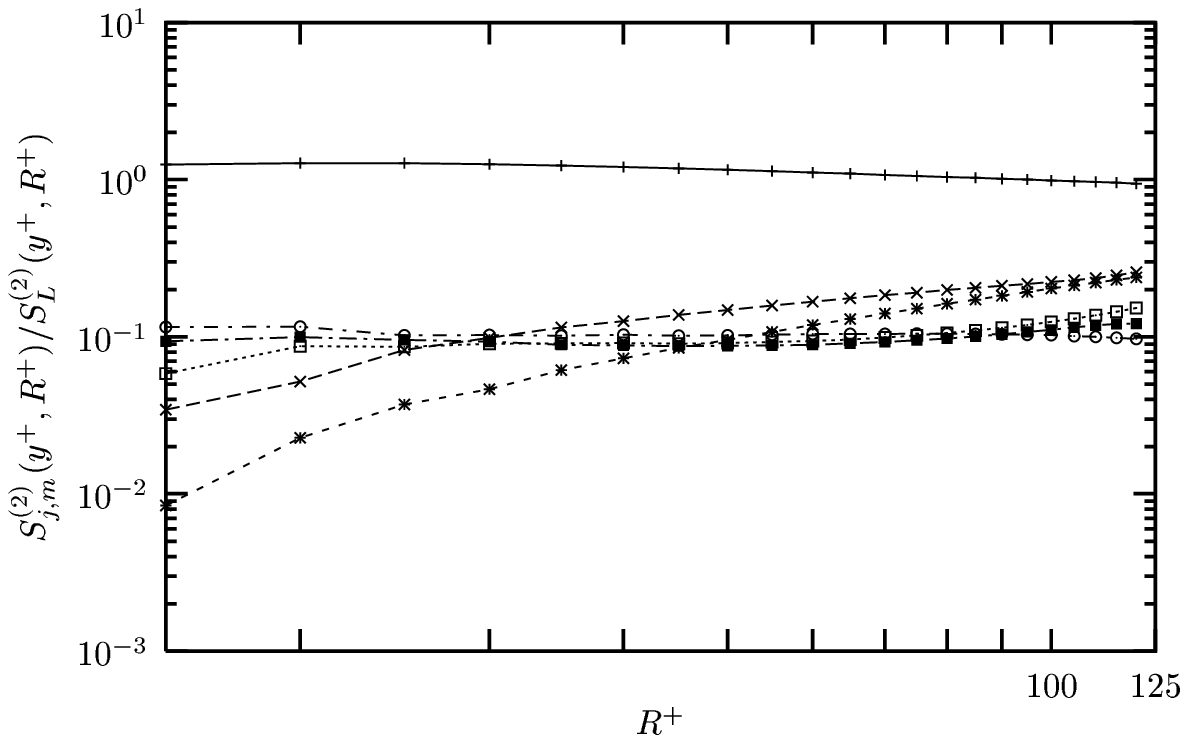}}}}}
\caption{Ratio of each single $(j,m)$ amplitude $S^{(2)}_{j,m}(y^+,\R)$ 
to the total structure function $S^{(2)}_L(y^+, \R)$ with 
${\bf R}$ in the direction of the mean flow and $y^+=62$. The 
$(j,m)$ indexes are: $(0,0)~(+)$; $(2,0)~(\times)$; 
$(2,2)~(\ast)$; $(4,0)~(\square)$; 
$(4,2)~(\blacksquare)$ and $(4,4)~(\circ)$.}
\label{fig7}
\end{figure}

\begin{figure}
\epsfxsize=1.2 \hsize
{\hskip -0.8cm{\centerline{{\epsfbox{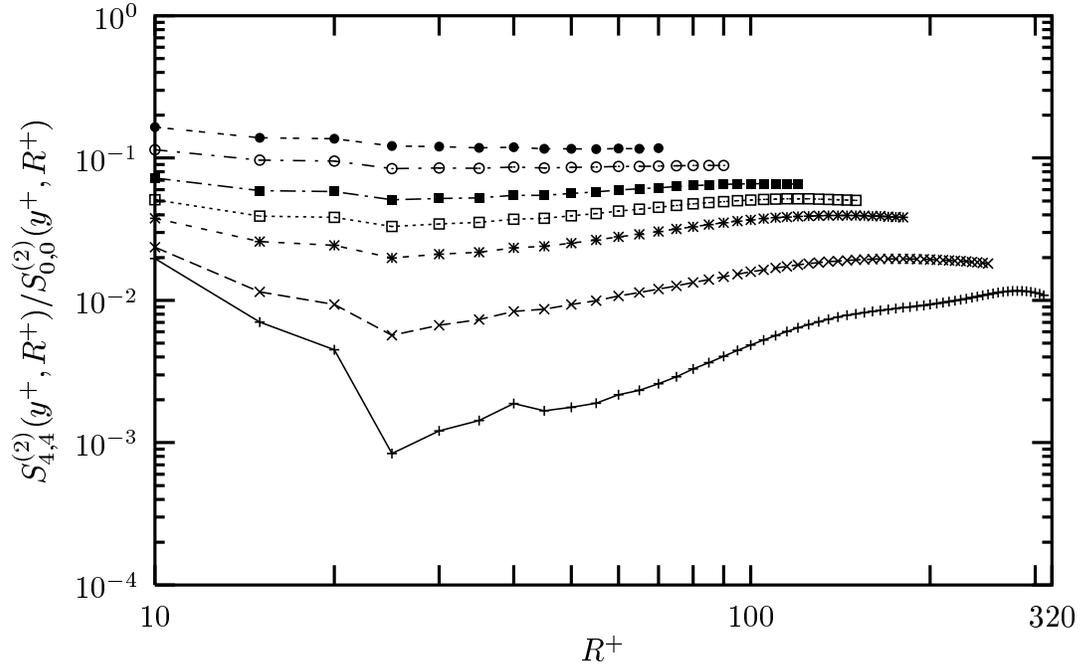}}}}}
\caption{Ratios between the sector $(j,m)=(4,4)$ of the decomposition of $S^{(2)}_L(y^+,R^+)$ and the isotropic sector $(j,m)=(0,0)$ as functions of R, at changing the analyzed height in the channel: $y^+= 160~(+)$; $y^+=125~(\times)$; $y^+= 92~(\ast)$; $
y^+= 80~(\square)$; $y^+= 62~(\blacksquare)$; $y^+= 48~(\circ)$ and $y^+= 37~(\bullet)$.}
\label{fig8}
\end{figure}

\begin{figure}
\epsfxsize=1.2 \hsize
{\hskip -0.8cm{\centerline{{\epsfbox{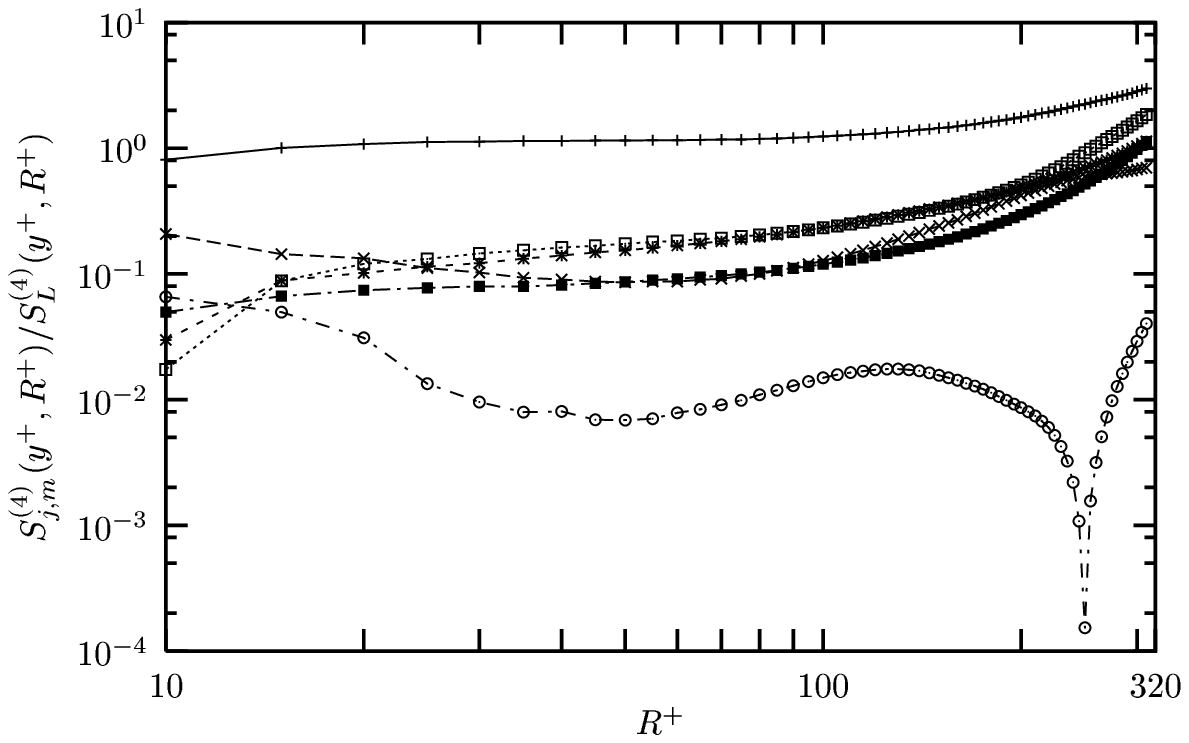}}}}}
\caption{Ratio of each single $(j,m)$ amplitude $S^{(4)}_{j,m}(y^+,\R)$ 
to the total structure function $S^{(4)}_L(y^+, \R)$ with ${\bf R}$ in
 the direction of the mean flow and $y^+=160$. The $(j,m)$ indexes
 are: $(0,0)~(+)$; $(2,0)~(\times)$; $(2,2)~(\ast)$; 
$(4,0)~(\square)$; $(4,2)~(\blacksquare)$ and $(4,4)~(\circ)$.}
\label{fig9}
\end{figure}

\begin{figure}
\epsfxsize=1.2\hsize{\hskip
 -0.8cm{\centerline{{\epsfbox{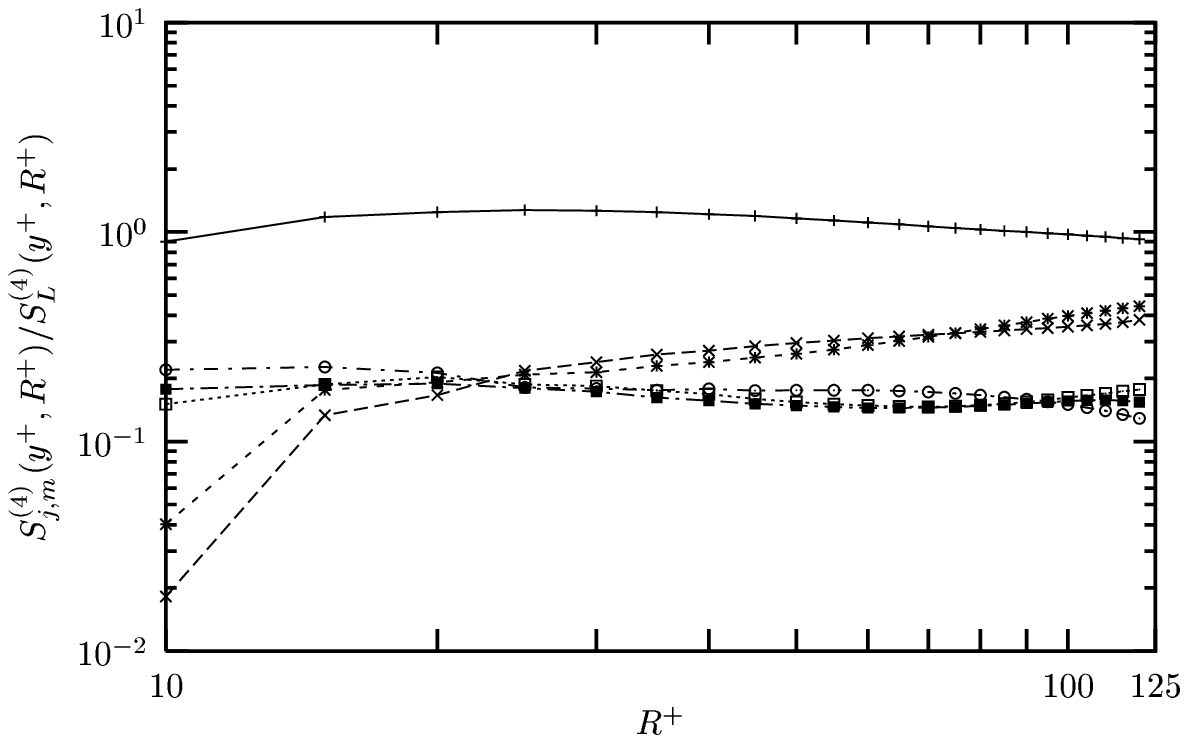}}}}}
\caption{Ratio of each single $(j,m)$ amplitude $S^{(4)}_{j,m}(y^+,\R)$ 
to the total structure function $S^{(4)}_L(y^+, \R)$ with ${\bf R}$ 
in the direction of the mean flow and $y^+=62$. The $(j,m)$ indexes 
are: $(0,0)~(+)$; $(2,0)~(\times)$; $(2,2)~(\ast)$; $(4,0)~(\square)$; 
$(4,2)~(\blacksquare)$ and $(4,4)~(\circ)$.}
\label{fig10}
\end{figure}

\begin{figure}
\epsfxsize=1.05\hsize{\hskip -0.2cm{\centerline{{\epsfbox{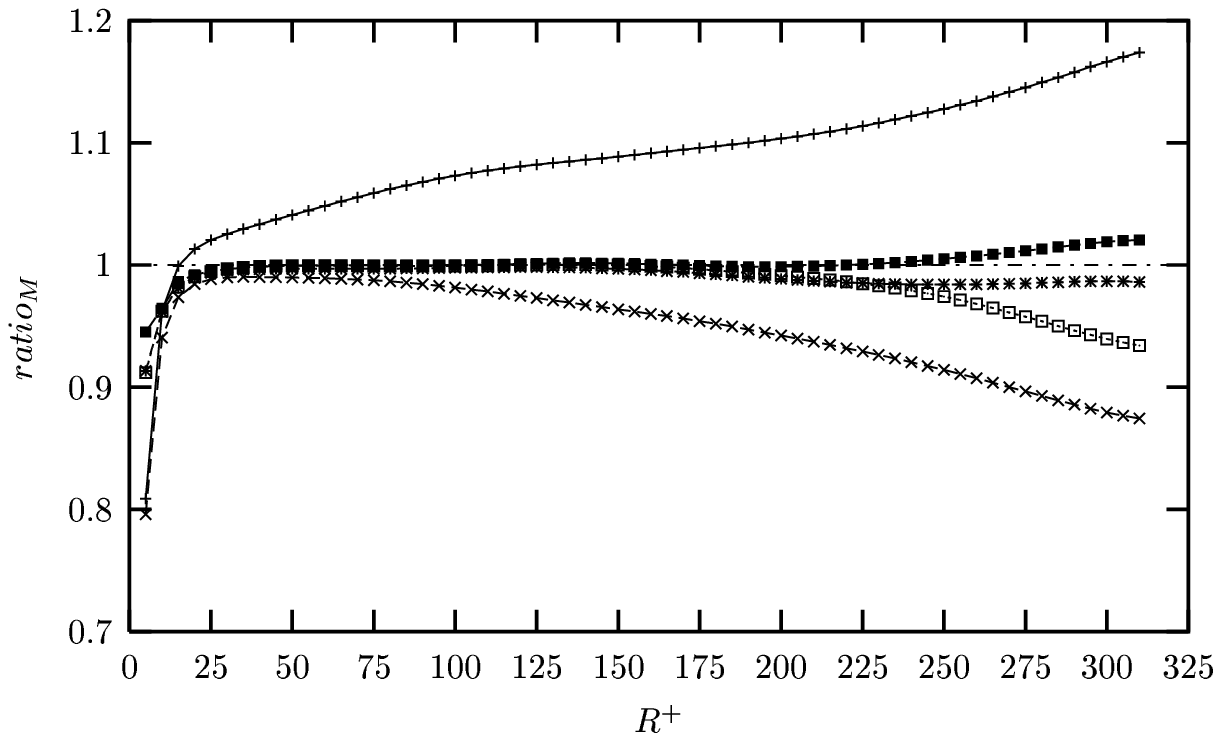}}}}}
\caption{Analysis of the convergence of the $SO(2)$ decomposition:
$ratio_{M}$ represents the ratio between the longitudinal 
structure function of order $2$
in the stream-wise direction reconstructed up to
 $M=0~(+)$; $M=2~(\times)$; $M=4~(\ast)$;
$M=6~(\square)$ and $M=8~(\blacksquare)$ and the 
undecomposed structure function, at the center of the channel $y^+=160$} 
\label{fig11}
\end{figure}

\begin{figure}
\epsfxsize=1.05\hsize{\hskip -0.2cm{\centerline{{\epsfbox{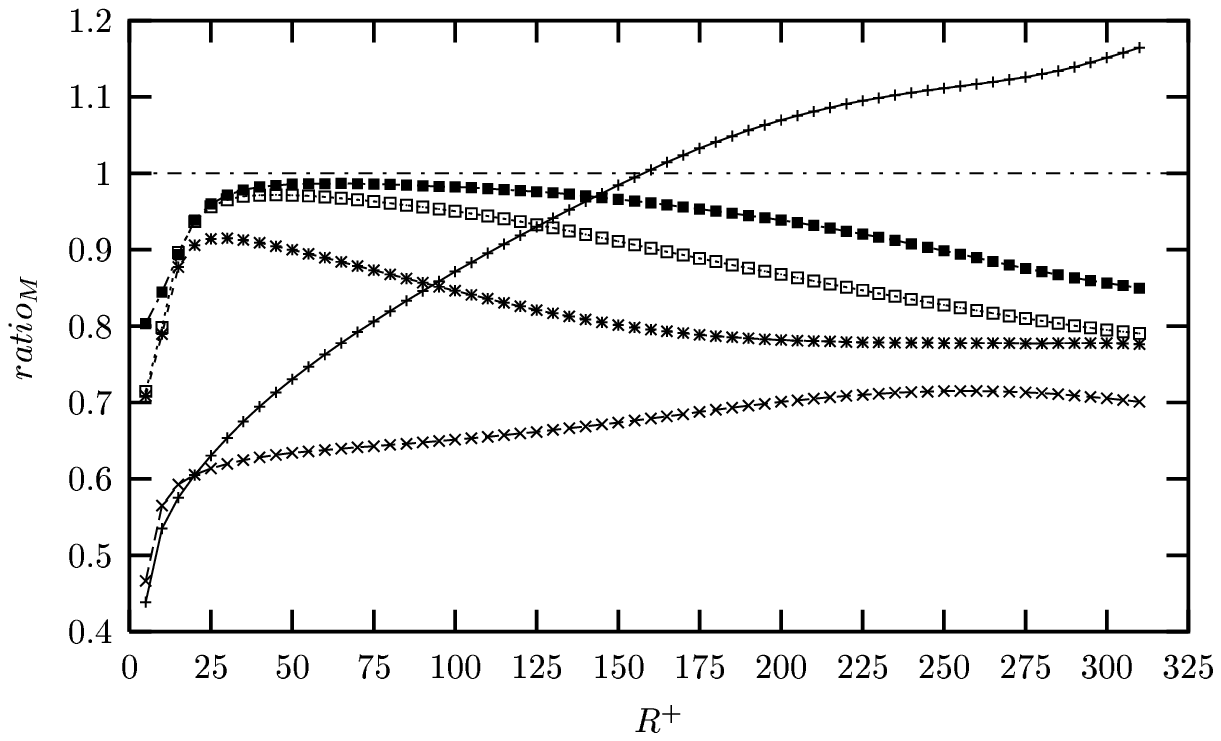}}}}}
\caption{Analysis of the convergence of the $SO(2)$ decomposition:
$ratio_{M}$ represents the ratio between the longitudinal structure function of order $2$
in the stream-wise direction reconstructed up to
$M=0~(+)$; $M=2~(\times)$; $M=4~(\ast)$;
$M=6~(\square)$ and $M=8~(\blacksquare)$ and the 
undecomposed structure function, at $y^+=37$} 
\label{fig12}
\end{figure}

\begin{figure}
\epsfxsize=1.05\hsize{\hskip -0.2cm{\centerline{{\epsfbox{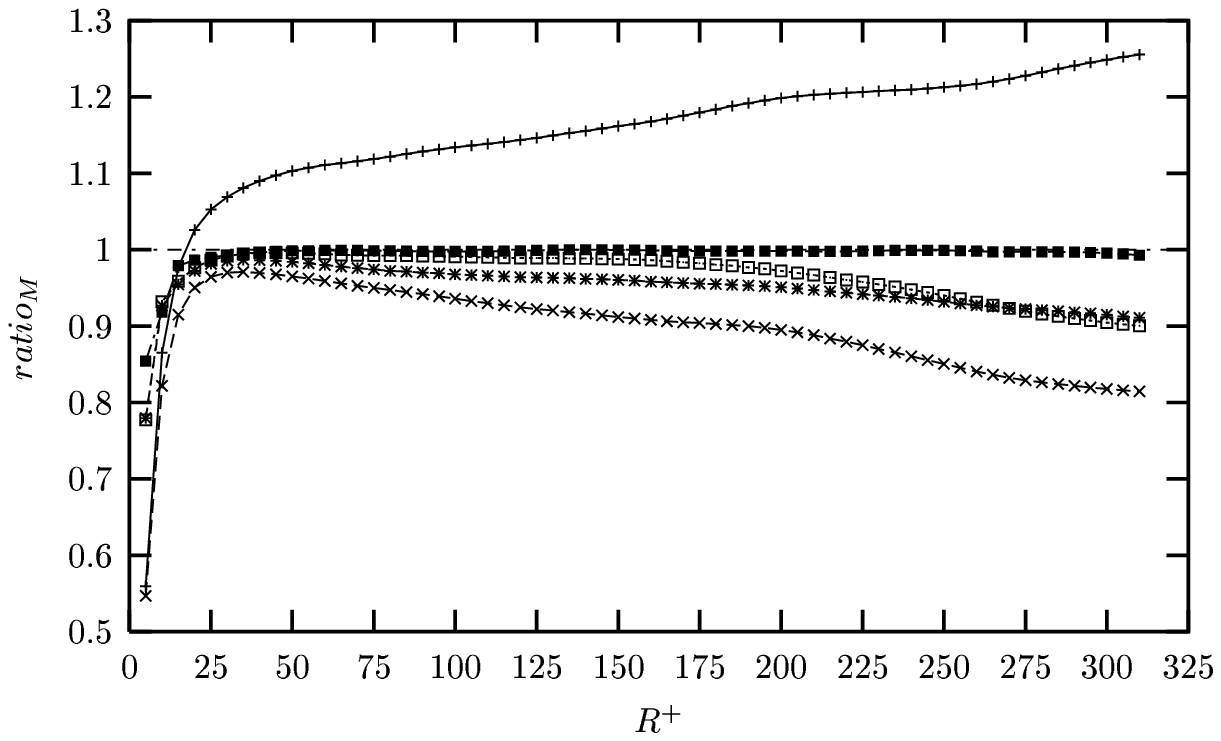}}}}}
\caption{Analysis of the convergence of the $SO(2)$ decomposition:
$ratio_{M}$ represents the ratio between the longitudinal structure function of order $4$
in the stream-wise direction reconstructed up to
$M=0~(+)$; $M=2~(\times)$; $M=4~(\ast)$;
$M=6~(\square)$ and $M=8~(\blacksquare)$ and the 
undecomposed structure function, at the center of the channel $y^+=160$} 
\label{fig13}
\end{figure}

\begin{figure}
\epsfxsize=1.05\hsize{\hskip -0.2cm{\centerline{{\epsfbox{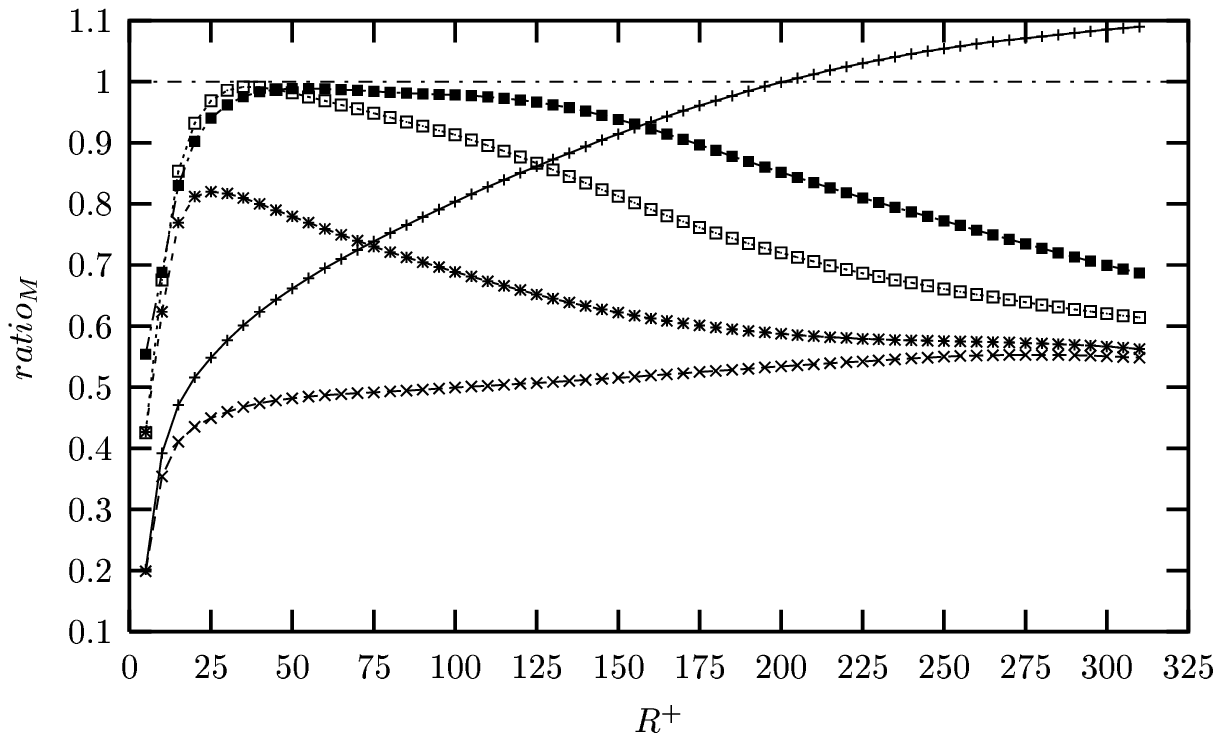}}}}}
\caption{Analysis of the convergence of the $SO(2)$ decomposition:
$ratio_{M}$ represents the ratio between the longitudinal structure function of order $4$
in the stream-wise direction reconstructed up to
 $M=0~(+)$; $M=2~(\times)$; $M=4~(\ast)$;
$M=6~(\square)$ and $M=8~(\blacksquare)$ and the 
undecomposed structure function, at $y^+=37$} 
\label{fig14}
\end{figure}

\begin{figure}
\epsfxsize=1.05\hsize{\hskip -0.2cm{\epsfbox{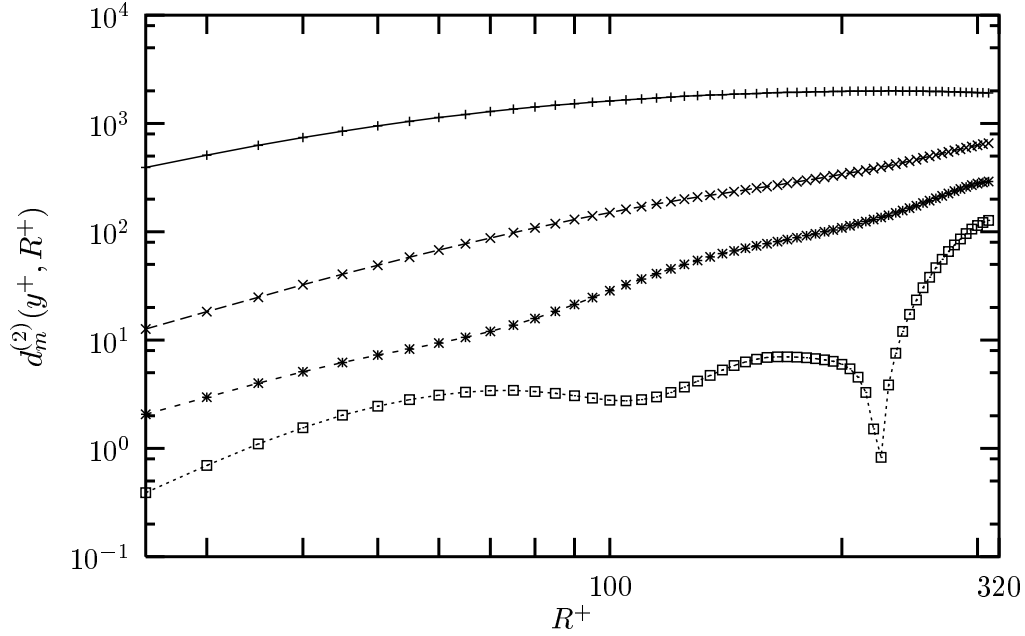}}}
\caption{Absolute weight of different $d_m^{(p)}(y^+,R^+)$ contributions 
for the second order structure function at the center of the channel $y^+=160$.
The $m$'s values of these components are: $0~(+)$; $2~(\times)$; 
$4~(\ast)$ 
and $6~(\square)$.}
\label{fig15}
\end{figure}

\begin{figure}
\epsfxsize=1.05\hsize{\hskip -0.2cm{\epsfbox{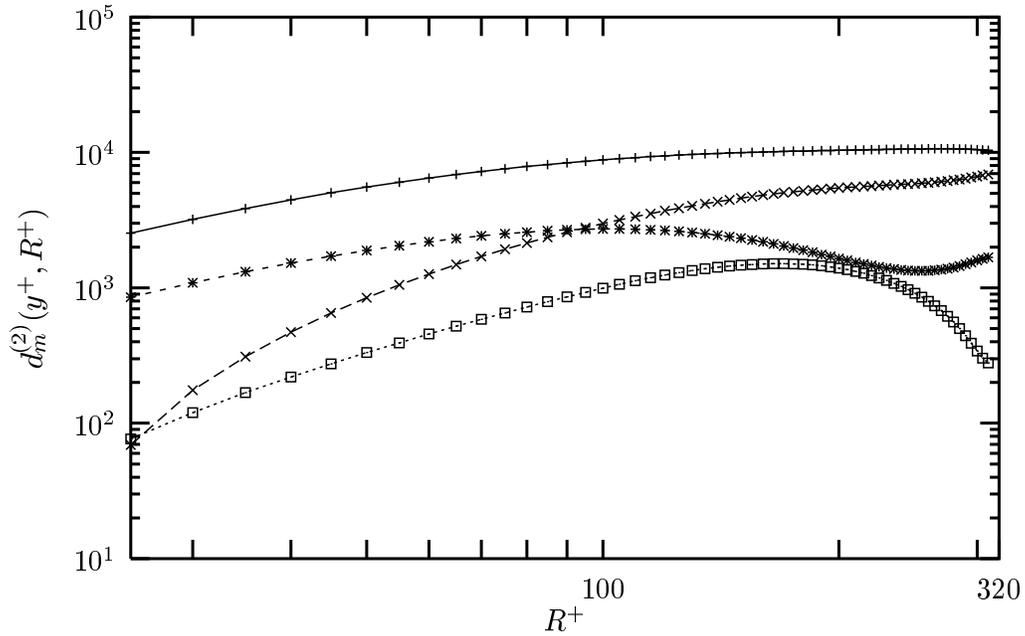}}}
\caption{Absolute weight of different $d_m^{(p)}(y^+,R^+)$ contributions 
for the second order structure function in the buffer layer  $y^+=37$.
The $m$'s values of these components are: $0~(+)$; $2~(\times)$; $4~(\ast)$ 
and $6~(\square)$.}
\label{fig16}
\end{figure}

\clearpage
\begin{figure}
\epsfxsize=1.1\hsize{\hskip -0.9cm{\epsfbox{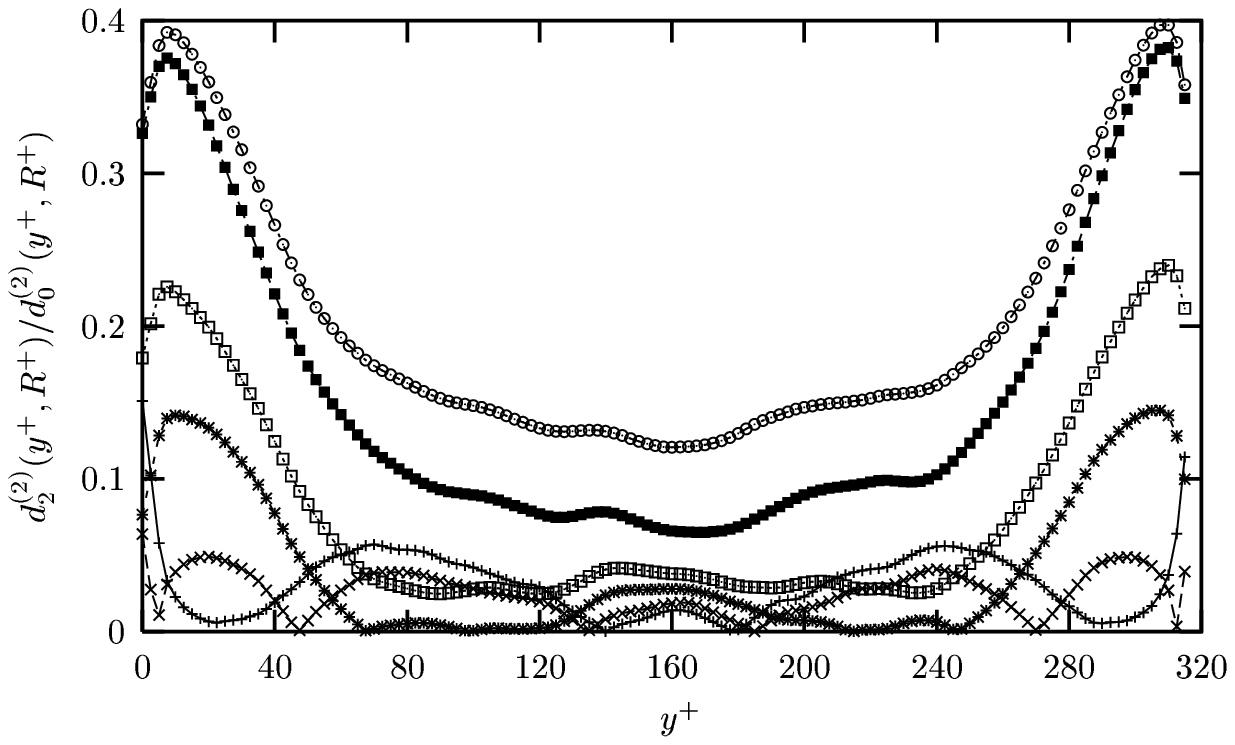}}}
\caption{Ratio between the projection on the $m=2$ sector and on the isotropic sector $m=0$ as a function of $y^+$, for $R^+=10~(+)$, $25~(\times)$, $50~(\ast)$, $75~(\square)$, $150~(\blacksquare)$ and $250~(\circ)$.}
\label{fig17a}
\end{figure}

\begin{figure}
\epsfxsize=1.1\hsize{\hskip -0.9cm{\epsfbox{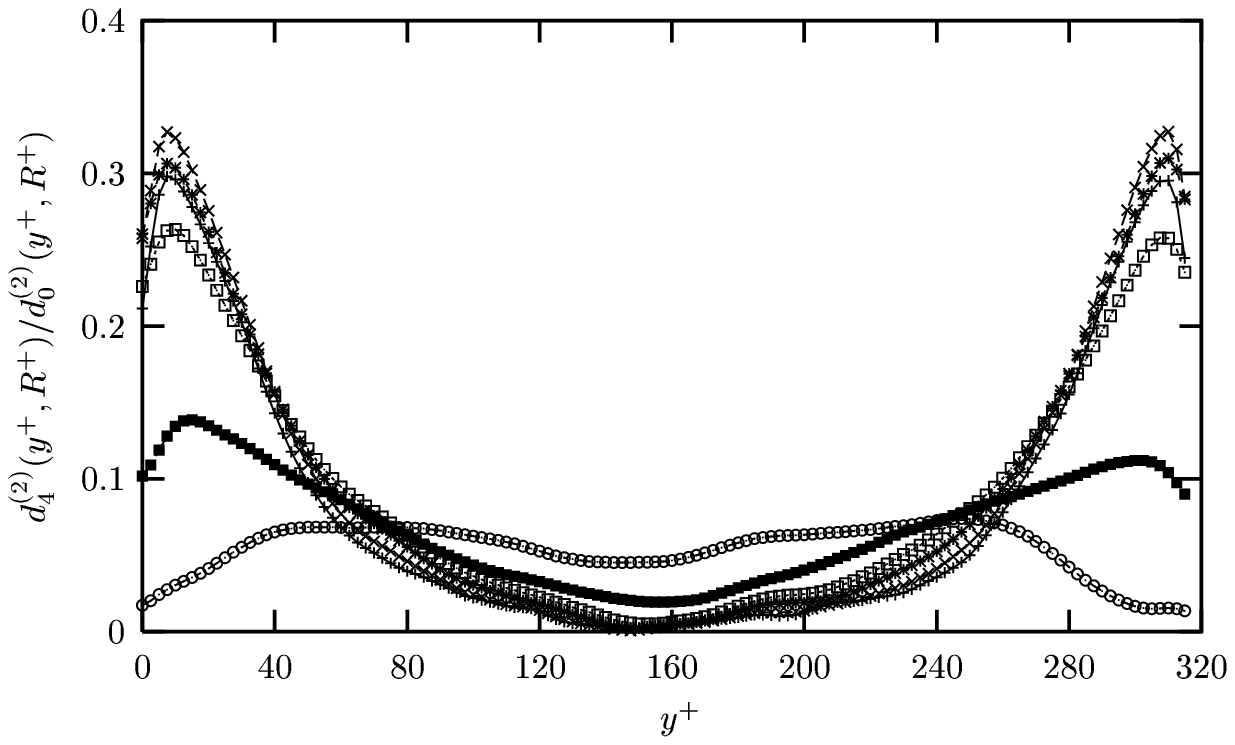}}}
\caption{Ratio between the projection on the $m=4$ sector and on the isotropic sector $m=0$ as a function of $y^+$, for $R^+=10~(+)$, $25~(\times)$, $50~(\ast)$, $75~(\square)$, $150~(\blacksquare)$ and $250~(\circ)$.
}
\label{fig17b}
\end{figure}

\begin{figure}
\epsfxsize=1.\hsize{\hskip 1.5cm\centerline{\epsfbox{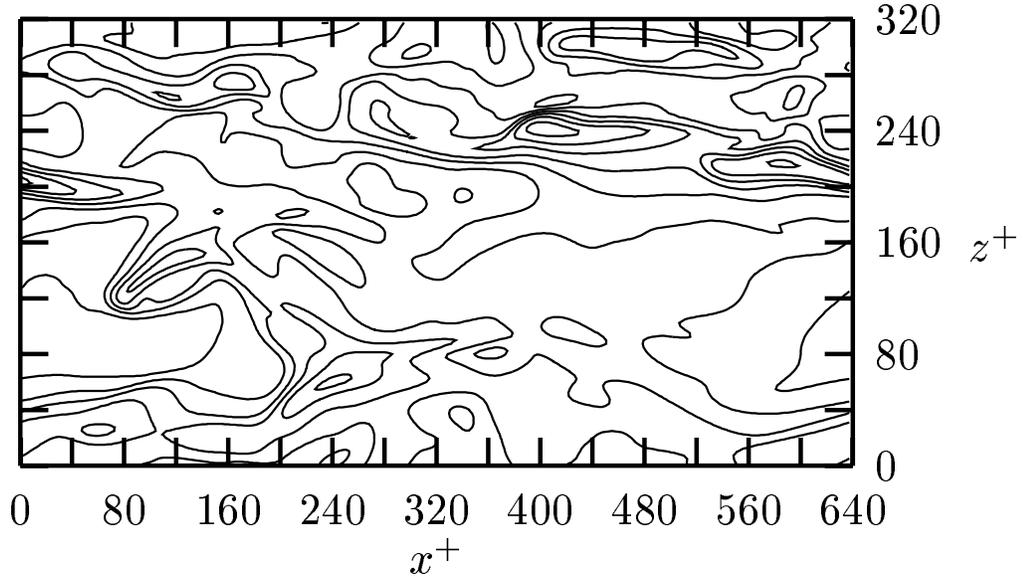}}}
\caption{Contour plot of the fluctuation of the stream-wise velocity at 
$y^+=37$.}
\label{fig17}
\end{figure}

\begin{figure}
\epsfxsize=1.2\hsize{\hskip -2.7cm{\epsfbox{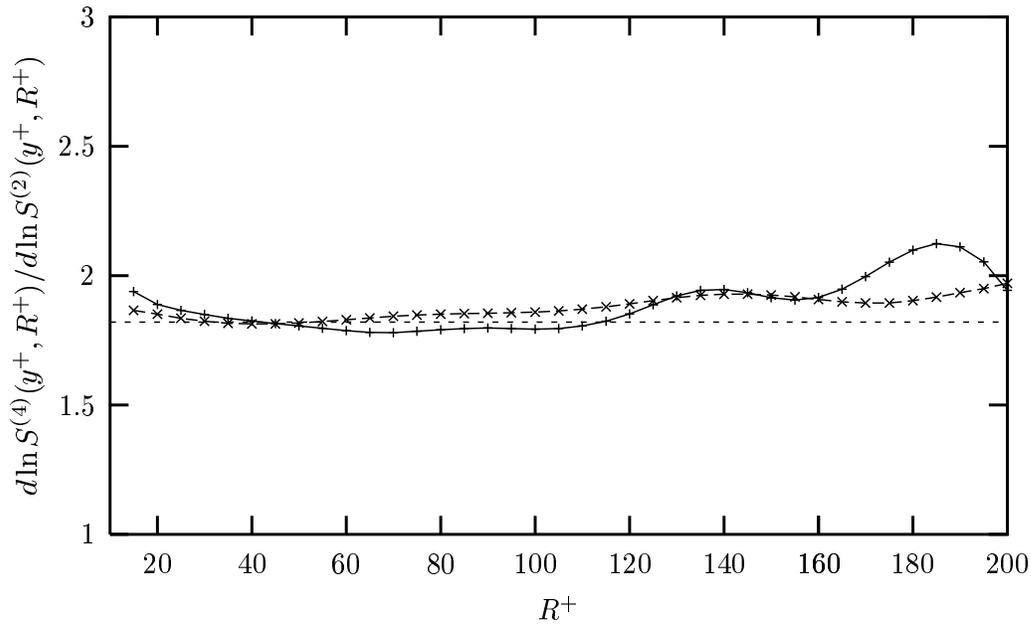}}}
\caption{ESS logarithmic local slopes of the undecomposed structure function 
in the stream-wise direction ($+$) and 
of the projection on the $m=0$ sector ($\times$) as functions of the 
scale $R^+$, for the moments $p=4$ versus $p=2$, at $y^+=160$.
The dashed line represents the value $1.84$ resulting from 
the experimental high-Reynolds numbers isotropic measurements}
\label{fig18}
\end{figure}

\begin{figure}
\epsfxsize=1.2\hsize{\hskip -2.7cm{\epsfbox{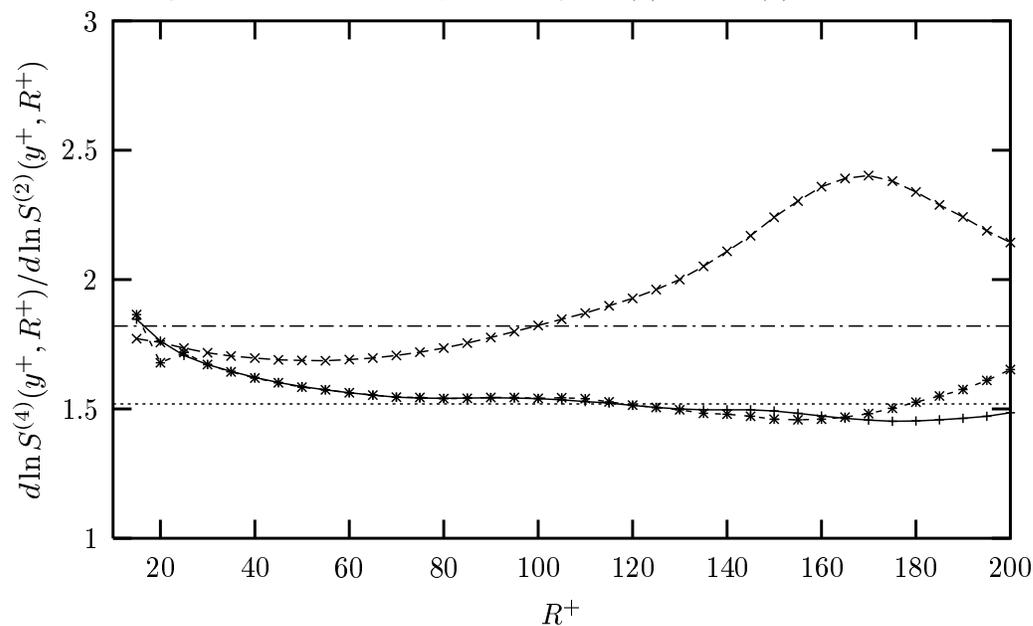}}}
\caption{ESS logarithmic local slopes of the undecomposed structure function 
in the stream-wise direction ($+$); 
of the projection on the $m=0$ sector ($\times$) and of the
 reconstruction 
up to $m_{max}=2$ ($\ast$) as functions of the scale $R^+$, 
for the moments $p=4$ versus 
$p=2$, at $y^+=37$.
The dotted-dashed line corresponds the best fit value, $1.52$,
 for the ESS logarithmic local slopes of the 
undecomposed structure function
in the stream-wise direction, the dotted line corresponds to the 
high-Reynolds number experimental isotropic value, $1.84$, for the same
quantity.}
\label{fig19}
\end{figure}
\end{document}